# Thermo-mechanical Properties of Hierarchical Biocomposite Materials from Photosynthetic Microorganisms

Israel Kellersztein[1,2,*], Daniel Tish[3], John Pederson[1], Martin Bechthold[3], and Chiara Daraio[1,*]

[1]Division of Engineering and Applied Science, California Institute of Technology, Pasadena, CA, USA.
[2]Resnick Sustainability Institute, California Institute of Technology, Pasadena, CA, USA.
[3]Material Processes and Systems Group, Graduate School of Design, Harvard University, Cambridge, MA, USA.
*Corresponding authors

**Corresponding authors:**
Israel Kellersztein: israelke@caltech.edu
Chiara Daraio: daraio@caltech.edu

**Abstract**

Extrusion 3D-printing of biopolymers and natural fiber-based biocomposites allows for the fabrication of complex structures, ranging from gels for healthcare applications to eco-friendly structural materials. However, traditional polymer extrusion demands high-energy consumption to pre-heat the slurries and reduce material viscosity. Additionally, natural fiber reinforcement often requires harsh treatments to improve adhesion to the matrix. Here, we overcome these challenges by introducing a systematic framework to fabricate natural biocomposite materials via a sustainable and scalable process. Using *Chlorella vulgaris* microalgae as the matrix, we optimize the bioink composition and the 3D-printing process to fabricate multifunctional, lightweight, hierarchical materials. A systematic dehydration approach prevents cracking and failure of the 3D-printed structure, maintaining a continuous morphology of aggregated microalgae cells that can withstand high shear forces during processing. Hydroxyethyl cellulose acts as a binder and reinforcement for *Chlorella* cells, leading to biocomposites with a bending stiffness above 1.5 GPa. The *Chlorella* biocomposites demonstrate isotropic heat transfer, functioning as effective thermal insulators with a thermal conductivity of 0.10 W/mK at room temperature. These materials show promise in applications requiring balanced thermal insulation and structural capabilities, positioning them as a sustainable alternative to conventional materials in response to increasing global materials demand.

## 1. Introduction

3D printing is a versatile processing method that facilitates the fabrication of biocomposites and biopolymers by enabling the production of complex geometries and customized designs with precise control over material deposition[1,2]. Extrusion 3D printing, which accommodates a wide range of biopolymer compositions, has been widely used in various fields including healthcare[3–5] and structural applications[6,7]. Biomaterials used in tissue regeneration[8–10] or environmental applications[11,12], for instance, are often gel-based, capable of delivering nutrients effectively to targeted cells and promoting cell proliferation[8,13,14]. These gels often exhibit soft mechanical properties, with Young's modulus values typically in the range of ~1 to 200 KPa, contingent on the crosslinking density of the printed gels and reinforcement components[15–17]. Fabricating scaffolds via 3D printing can involve the use of stiffer materials, such as poly (lactic acid) and polycaprolactone, with mechanical properties tailored to match the stiffness of the target tissue for regeneration, e.g., bone, reaching a Young's modulus ranging from 40 MPa to 1 GPa[18–21].

Polymer biocomposites for structural applications, typically composed of a polymer matrix reinforced with wood or natural fibers, are intended to withstand larger stresses, necessitating higher stiffness. Similar to conventional processing methods like injection molding or extrusion, 3D printing these biocomposites involves significant fabrication challenges. The polymer matrices often exhibit high viscosities, complicating conventional extrusion 3D printing unless high temperatures are involved, resulting in high energy consumption[22–24]. Moreover, optimizing the interaction between the matrix and the reinforcement materials requires meticulous treatments to enhance stress transfer efficiency [25–29].

Microalgae offer promising solutions to these challenges. These aquatic, photosynthetic microorganisms are globally abundant and adaptable to diverse environments for growing, including seawater, freshwater, and bioreactors[30,31]. Microalgae, with their diverse composition of proteins, lipids, and polysaccharide, as well as their unique cell morphology, have become central to sustainable research across various applications, including biofuels, food additives, cosmetics and wastewater purification[32–36]. Recent advancements in materials science and engineering have demonstrated the use of microalgae biomass as a natural polymer source for bioplastic synthesis[37–39]. Additionally, microalgae have been explored as fillers in polymer blends[40,41], composites[42–44], and cement[45] through conventional fabrication processes, such as compounding followed by injection molding[42], compression molding [41,43,44], and solvent casting[40]. These studies report a decrease in the mechanical properties of the blends and biocomposites with increasing microalgae concentration, for example, *Chlorella* reduced the strength and stiffness of polyethylene by 60% and 50% respectively, whereas the strength and stiffness of poly(lactic acid) were reduced by 61% and 40% respectively, when introducing microalgae into the material composition[40–44]. Additionally, research indicates that adding microalgae to cement reduces the matrix strength by up to ~80%[45].

Recent studies have focused on adapting microalgae for extrusion 3D printing, offering a sustainable and renewable material source[46,47]. 3D printing with microalgae presents a more sustainable alternative to conventional processing methods, such as printing at room temperature and using water as a solvent to control bioink rheology. *Chlorella vulgaris* microalgae suspensions, with varying particle volume fractions (up to 60 vol.%), were produced and non-polar oils were incorporated into the suspensions to minimize macroscopic defects in printed structures post-dehydration. Although the mechanical properties of the printed *Chlorella*-based materials were not investigated, the evaporation of water and oil during dehydration introduced defects within the materials' microstructure[46], likely impacting the mechanical properties. In a subsequent study, the reinforcement effect of cellulose fibers and drying methods, such as freeze-drying, oven-drying and desiccator, on the mechanical properties of 3D printed *Spirulina* biocomposites were investigated[47]. At the macroscale, they observed that freeze-drying the materials maintained their integrity, while oven and desiccator drying caused cracks and deformation[47]. At the microscale, they reported that oven and desiccator drying introduced microcracks, while freeze-drying resulted in a foam-like structure from solvent evaporation[47]. Cellulose fibers were required to deliver a strong material, with a large concentration (20 wt.%) of fibers leading to a compression strength of 16.4 MPa, though the modulus did not exhibit a similar improvement trend[47]. All these prior studies demonstrate that 3D printing offers flexibility in designing structures and achieving complex geometrical shapes[48], however, controlling post-printing processes, such as dehydration of natural-based bioink, is crucial for maximizing the mechanical functionality of the material.

Building on prior examples[46,47], in this study, we improve fabrication approaches and post-printing processes to obtain lightweight, hierarchical *Chlorella*-based biocomposite materials using extrusion 3D printing (Fig. 1), without the introduction of any petrochemical components. Our methodology optimizes biocomposite ink formulation, including 2-Hydroxyethyl Cellulose (HEC) reinforcement, and material dehydration to achieve final biocomposite materials with superior mechanical and thermal properties. We control the rheology of the slurries to improve ink flow and overall printability. We detail the role of processing parameters and reinforcement concentration on the mechanical and thermal properties of the *Chlorella* biocomposites. The final materials show a record compression strength of 23 MPa and a bending stiffness of 1.6 GPa, and a low thermal conductivity of 0.1 W/mK at room temperature. These properties are comparable to commodity plastics, like polyolefins and vinyls, and wood-based composites. Our approach is adaptable to other microalgae systems, enabling the fabrication of intricate 3D structures on a large scale.

## 2. Bioink composition and 3D printing

The bioinks used in our work were formulated to minimize the embodied carbon and include no petrochemical compounds. Additionally, the bioinks were printed at room temperature and printing patterns were shaped precisely to reduce the processing waste.

The bioink composition integrates *Chlorella vulgaris* cells that have a spherical shape and are reported to have diameters ranging from 2 to 10 $\mu$m[49,50]. We used ultrapure *Chlorella* with a cell diameter of 3.21±0.21 $\mu$m in its dry state (Fig. S1). These specific microalgae were grown in glass tubes, diminishing the presence of additional components, or impurities, which may interact weakly with *Chlorella* cells, resulting in discontinuities and defects that could affect the final properties of the biocomposite material.

Microalgae-based biocomposites were 3D printed using viscous suspensions of *Chlorella* in water with HEC. The workflow includes the initial dispersion of HEC in water, and then, addition of the microalgae cells into the mixture (Fig. 1I). HEC was integrated into the bioink to fulfill two objectives: (i) As a binder, enhancing cell immobilization through gelation[51]; (ii) as the reinforcement component in the biocomposite, as HEC can create strong physical interactions, such as hydrogen bonding, with the *Chlorella* cells[52]. The bioink was printed into mechanically stable hierarchical structures of cubic and rectangular shapes (Fig. 1II-IV).

To define the optimal water content in the bioink, we selected an initial HEC concentration of 5 wt.%, relative to the biomass concentration (Fig. 2A). At water concentrations below 56 wt.%, the bioink was too viscous to be printed, while at water concentrations above 66 wt.%, the bioink was too fluid to be precisely deposited using the minimum pressure of the 3D printer. After dehydrating the printed samples using the protocol described in section 3.2, we observed an inverse relationship between volumetric shrinkage and density: higher water concentrations in the bioink led to increased evaporation, causing greater shrinkage and reduced density.

After printing, inconsistencies in sample weight were observed despite using consistent printing conditions (Fig. S2a). These variations, which can affect the density and mechanical properties of the biocomposites, were attributed to the increased viscosity of the *Chlorella*-based bioink over time (Fig. S2b-c). HEC is commonly used as a rheological additive to improve the printability and shape fidelity of bioinks for extrusion 3D printing by increasing viscosity and enhancing flow behavior[53]. This increase in viscosity is likely the result of HEC's physical gelation, resulting from physical interactions, e.g., hydrogen bonding and van der Waals forces, between the HEC macromolecules and water[54]. To ensure the consistency of 3D printed parts, the mass flow rate principle (discussed in supplementary note #1) was applied to adjust the printing speed for every 3D printed sample. This approach ensured that the same amount of material was deposited during extrusion 3D printing, resulting in structures with similar weight, regardless of the different printing parameters (Fig. S2d).

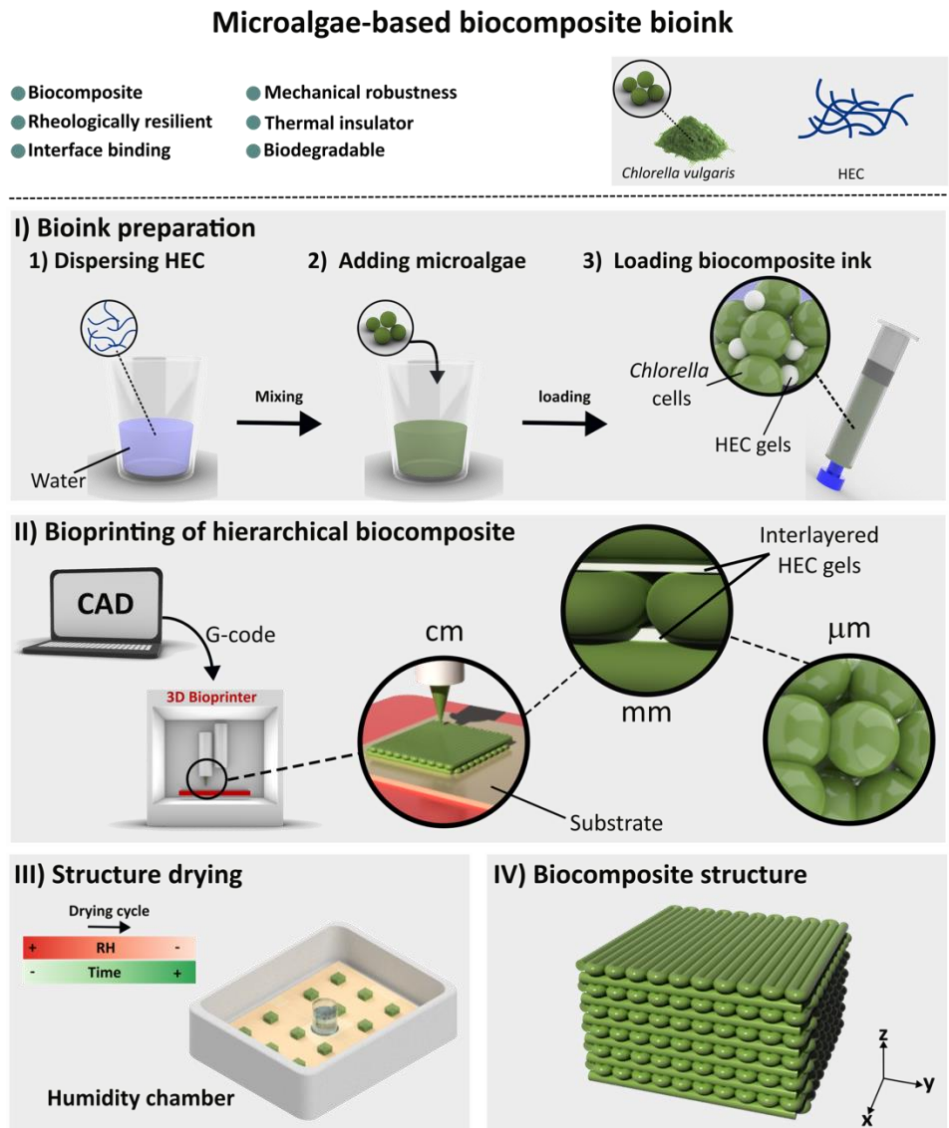

**Fig.1 Schematic diagram of the fabrication processes used in *Chlorella vulgaris* biocomposites.** The top row lists the properties and components used. Panels (I-IV) describe the steps for the fabrication of *Chlorella* biocomposites, highlighting the micro-, milli-, and centimeter length scales within the material's hierarchical structure. The final samples fabricated are 20 x 20 x 10 mm$^3$.

## 3. Results

We studied multiple printing and dehydration parameters, to maximize the materials' properties of the printed biocomposites. This included a detailed characterization of the bioink rheology and selection of the ink's composition, to improve printability (Fig. 2A-L). After fabrication and dehydration, we performed structural, mechanical (Fig. 3A-L), and thermal analysis (Fig. 4A-D) of the final biocomposites. The parameters' selection followed an iterative process that evaluated the role of each variable as a function of all the others.

### 3.1 Shape integrity maintained via bioink thixotropic behavior

In particle suspensions, interactions among different cells can significantly affect the rheological behavior of the bioink. High particle concentrations can form a continuous and interconnected network that resists flow, and high shear forces are required to start the flowing process through the printer nozzle[55,56]. *Chlorella*-based bioink exhibited key characteristics for successful printing: it behaved as a viscoelastic solid ($G' > G''$) at low shear stresses (Fig. 2C), had a yield point before flowing (Fig. 2C-D), and demonstrated shear-thinning behavior with increased shear rate (Fig. 2E). Higher concentrations of HC generated additional gelation sites, promoting cell-gel interactions, and increasing both the shear stress at yield, measured from the $G'$ and $G''$ cross-over during the oscillation test, and viscosity of the bioink (Fig. 2C-E). The bioink solvent (water) had a more significant effect on the rheological properties compared to HEC (Fig. S4), as higher water concentrations reduced the shear stress at yield and the viscosity of the bioink. This is likely because of the reduction in the overall particle density in the bioink, thus reducing the overall cell-cell interactions[55,57,58]. A consistent water concentration of 58 wt.% was used for all bioinks from this point. This concentration allowed good printability, moderate volumetric shrinkage, relatively low densities (<1 g/cm³), and faster and controllable dehydration of the printed biocomposite.

The shear-thinning behavior of *Chlorella* biocomposite inks demonstrated reversible characteristics, with viscosity nearly returning to its original state after shear forces were removed (Fig. 2F), demonstrating thixotropic behavior. Under high shear rates, particularly influenced by the conical nozzle geometry, interactions between *Chlorella* particles were disrupted, allowing them to flow independently[59]. After extrusion through the nozzle, when shear forces diminished, cells reorganized and re-established cell-cell interactions, thereby restoring bioink viscosity and preserving the shape of 3D printed structures[55,59]. Despite *Chlorella* having a relatively thin cell wall (~12-14 nm) compared to other microalgae species, its fibrillar morphology and high extracellular polysaccharide content provided sufficient stiffness to maintain the structural integrity of the cell under harsh flow conditions[49,50]. This resilience was critical as high shear forces deform the cells, impacting the rheological properties of the bioink.

Deep evaluation of the effect of HEC on the rheological properties of the bioink yielded an optimized 3D printing process, where a continuous and even deposition of biocomposite filaments during fabrication was achieved. We noticed that increasing the HEC concentrations within the bioink slightly reduced the filament diameter (Fig. 2G), however, no infill pattern detachment was observed after dehydration using consistent infill distance while 3D printing different HEC concentrations.

**3.2 Slow dehydration rates diminishes differential shrinkage and structure failure after 3D printing**

Dehydration plays an important role in the final mechanical properties of the biocomposites and specifically for *Chlorella*-based materials, it is influenced by the initial water concentration in the bioink. Water concentration affects the volumetric shrinkage and density of the printed materials, along with the bioink printing pressure (Fig. 2A). Regardless of the printing parameters used, when samples were dried at ambient conditions, or in an oven, we observed that the morphology of the biocomposites exhibited numerous defects (Fig. 2B-I), often leading to catastrophic failure, in agreement to previous work[46,47,60].

To address this, we developed a controlled, multiphase dehydration protocol (Fig. S3a), inspired by the sol-gel theory of drying[61] (see supplementary note #2). The first stage of drying, known as the constant rate period (CRP), is where most of the structure's shrinkage occurs (Fig. 3G), maximizing internal drying stresses[62]. These internal tensile stresses are primarily concentrated near the drying surface, causing differential shrinkage of the structure. When these internal stresses are higher than the strength of the material, cracking and eventually catastrophic failure of the 3D printed part occurs. To mitigate differential shrinkage in the 3D printed biocompoiste, the CRP needs to be slow. Following 3D printing, the wet structures were placed in an environmental chamber at a relative humidity (RH) of 75% for two days (see experimental section in supporting information). At the beginning, the RH in the chamber increased as water is being rapidly evaporated (Fig. S3b-c). After two days, when shrinking stabilized, the samples were moved into a humidity chamber at a RH~50% for four days. During this second stage of drying, known as falling rate period, evaporation occurs at a lower rate[63], leading to successful, and crack-free, dehydrated 3D printed structures.

## 3.3 Higher HEC concentration reduces shrinkage and enhances *Chlorella* cells adhesion, promoting a continuous micromorphology

The HEC concentrations utilized in this study were carefully selected to improve the printing process and ensure the structural integrity of the printed structures. A 3 wt.% HEC concentration was determined to be the minimum threshold necessary to produce dehydrated samples without deformation or cracking. Conversely, 10 wt.% represented the maximum printable concentration, as higher HEC levels significantly increased viscosity. Exceeding this concentration resulted in a prolonged 3D printing process using the maximum pressure of our 3D printer, during which the samples began to dry at room temperature, ultimately leading to warping and delamination of the printed layers. It is important to note that, in the absence of HEC, printed structures catastrophically failed during dehydration because of (i) the high differential shrinkage developed within the material during the dehydration process; and (ii) the absence of binding effect provided by HEC gelation.

To maintain the 3D printed geometry as consistent as possible after printing and dehydration, it is essential to reduce volumetric shrinkage. For biocomposites made with *Chlorella* reinforced with HEC, the in-plane shrinkage is controlled by (i) the symmetry of the printing pattern (0°-90°, Fig. 2L); and (ii) the HEC concentration, whereas the out-of-plan shrinkage is controlled only by the HEC concentration (Fig. 2H). Our results show that increasing the HEC concentration from 3 wt.% to 10 wt.% reduced the linear shrinkage in the *x*-axis by 9%, in the *y*-axis by 18%, and in the *z*-axis by 21%, respectively. These results are consistent with prior work, which reported that increasing the viscosity of the bioink limited the shrinkage and deformation of the printed structures during dehydration[60].

The density of the biocomposite is also impacted by the HEC concentration (Fig. 2I). We observed that an increase in HEC concentration leads to proportionally higher density. Variations in density are correlated to the presence of microstructural defects, arising from the binding effect of HEC. At low HEC concentrations, the microstructure of the biocomposites presents larger defects, due to the concentration of trapped air within the structure, resulting in low density values (0.87±0.02 g/cm$^3$ for 3 wt.% HEC). Rising the HEC concentration increased the biocomposites density to 0.93±0.02 g/cm$^3$ at 5 wt.% HEC and 0.96±0.02 g/cm$^3$, for 10 wt.% HEC.

HEC binding resulted in a continuous matrix of aggregated cells, leading to a uniform microstructural morphology with a characteristic hierarchical organization. The structural organization of *Chlorella* biocomposites spanned nine hierarchical levels. At the molecular level, (i-ii) the key macromolecules in *Chlorella* cells included amino acids, which polymerized to form proteins; sugars, which assembled into polysaccharides, such as cellulose; and fatty acids, which combined to create lipids including triglycerides and phospholipids[49]. At the cellular level, these biopolymers aggregated into (iii) layers that form (iv) the cell walls[64]. These cell walls surround and protect the cell's internal components, including the chloroplast, nucleolus, and mitochondria, resulting in a (v) single cell at the whole-cell level. At the microlevel (iv), *Chlorella* cells were organized into a continuous particle aggregation morphology without visible microcracks under scanning electron microscope (SEM) (Fig. 2J). At the millimeter level, (vii) the particle aggregates were extruded as a filament and further arranged into (viii) a single layer. Post-printing, flow-induced orientation of HEC layers between microalgae was observed (Fig. 2K), showing the presence of HEC between printed layers. At the macrolevel (Fig. 2L), we achieved full control over printability and dehydration, resulting in continuous, crack-free, (ix) 3D printed samples. However, structures with only 3 wt.% HEC randomly deformed during drying. This failure emphasizes that lower HEC concentrations are insufficient to effectively bind all biomass, necessitating optimization of HEC content for structural stability.

Our fabrication approach grants us control over multiple hierarchical levels within the *Chlorella* biocomposites structure, ensuring optimal performance across scales. At the microscale, we carefully manage the distribution and concentration of both *Chlorella* cells and HEC within the bioink, directly influencing the material's properties and structural integrity. This control extends to the millimeter scale, where we optimize the arrangement and interaction of biocomposite layers, where flow-induced HEC interlayer positioning enhances cohesion and consistency throughout the material. This alternating layer organization significantly impacts meso-level structure by strengthening interlayer bonds, which is critical for mechanical strength. At the macroscale, we design the overall architecture of the printed structures to align with specific functional goals. By systematically refining each stage of the fabrication process—from bioink formulation to final structural design—we ensure that the controlled hierarchical levels contribute effectively to the biocomposite's overall performance.

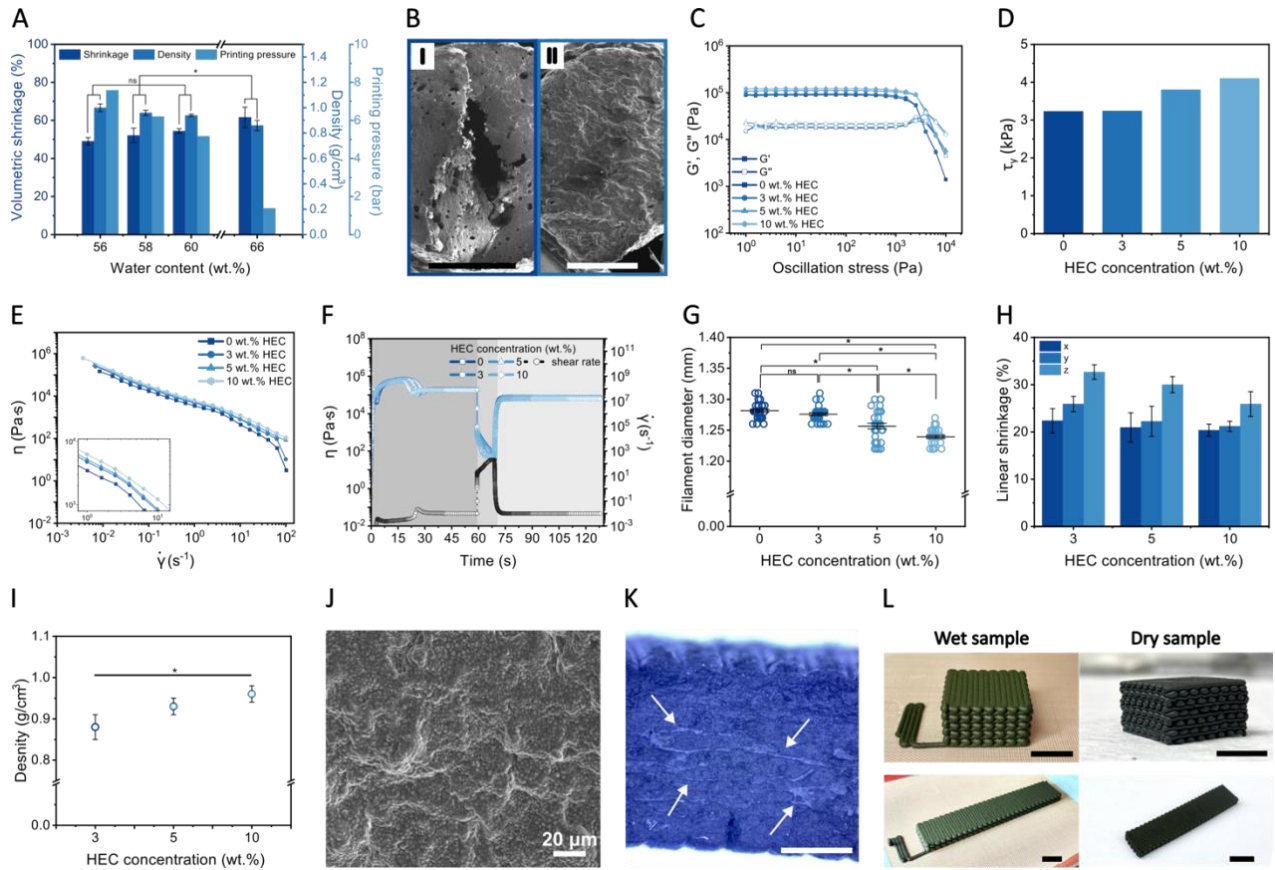

**Fig. 2** *Printability and hierarchical morphology of the biocomposite*. (A) Volumetric shrinkage and density variation of the biocomposites after printing and drying, as a function of water concentration in the bioink. (B) Binding effect of HEC on the microstructure of *Chlorella* after 3D printing; (I) without HEC and (II) with 5 wt.% HEC; scale bars 500 μm. (C-D) Rheological properties of *Chlorella* biocomposites with higher HEC concentration. (E) *Chlorella* bioinks have a shear thinning behavior. (F) Fast viscosity recovery and shape retention of *Chlorella*-HEC bioinks. (G) Printed filament diameter decreases with HEC concentration. (H) Linear shrinkage decreases with HEC concentration. (I) The density of the composites increases with HEC concentration. (J-L) Hierarchical structure of biocomposites reinforced with 10 wt.% HEC. (J) *Chlorella* cells retain their shape after 3D printing and dehydration. (K) Morphological control of HEC gels between microalgae layers (pointed with white arrows); scale bar 4 mm. (L) Wet and dry structures with different aspect ratios showing the symmetric printing pattern (0°-90°); scale bars 1 cm. Group comparisons were performed by one-way ANOVA where ns denotes $p>0.05$ and * represents $p<0.05$.

### 3.4 Improved strength and stiffness of the biocomposite is achieved through HEC binding and reinforcing mechanisms

We first characterized the uniaxial mechanical performance of the biocomposites for different 3D printing parameters (i.e., nozzle diameter and layer height, Fig. S5) and bioink water concentration (Fig. S6). We optimize both the 3D printing process and the bioink formulation to maximize mechanical functionality. We then conducted quasi-static compression and 3-point bending tests to elucidate the reinforcement effect of HEC on the mechanical behavior of the dehydrated *Chlorella* biocomposites.

Nozzle diameter is an important parameter because it affects both the printing pressure, and the amount of material extruded during the printing process. We study the effect of systematic variations of nozzle diameter, increasing it from 0.84 mm to 1.6 mm. We observed that the compressive modulus of the biocomposite increased linearly from 80±4 MPa to 96±2 MPa (Fig. S5a). This can be qualitatively explained by the increased layer thickness deposited with larger nozzle diameters, resulting in denser filaments. Smaller nozzle diameters were also investigated; however, the printing time was excessively long, causing the biocompoiste to dry during 3D printing, resulting in warping and delamination of the structure. To study the role of layer heights, we fabricated biocomposite structures using a 1.19 mm nozzle diameter increasing the layer height from 0.9 mm to 1 mm (Fig. S5b). In this small range, we observed an increase in the compression modulus from 78±8 MPa to 115±5 MPa. This increase can be attributed to improved interlayer adhesion and reduced anisotropy of the overall structure, resulting in a denser structure with greater resistance to compressive forces. When layer heights above 1 mm were explored, a decrease in the modulus was observed, likely because of weaker interlayer adhesion and material flow issues. The water concentration influences both the final algae cell concentration in the bioink and the flow and dehydration process. However, the specific modulus ($E/\rho$) and the specific compressive strength ($\sigma/\rho$) of the biocomposites were not affected by changes in water concentration (Fig. S6).

After optimizing 3D printing processing and bioink optimization, we turned our attention to understanding the reinforcing effect of HEC (Fig. 3). Under uniaxial compression, all stress-strain curves of the biocomposites with different HEC concentration showed an initial linear regime (Fig. 3A). A progressive failure until break was observed after reaching the maximum strength of the materials. An increase of ~342% for the specific compressive modulus, which was calculated from the initial linear elastic part of the compression experiments (Fig. 3B) was observed when increasing the HEC concentration from 3 wt.% to 10 wt.%. Similarly, the specific strength of the material was enhanced by ~ 174% when the HEC concentration raised from 3 wt.% to 10 wt.% (Fig. 3C).

The force-displacement curves of the biocomposites, with different HEC concentrations, under 3-point bending exhibited a linear elastic regime during loading, followed by a catastrophic failure at maximum force, revealing a brittle fracture (Fig. 3D). *Chlorella* reinforced with 10 wt.% HEC presented a specific bending modulus ($E_f/\rho$) of 1.7 GPa・cm$^3$/g, 173% higher than the bending modulus obtained with a minimal HEC reinforcing concentration of 3 wt.% (Fig. 3E). Similarly, a maximum specific bending strength ($\sigma_f/\rho$) of 14 MPa・cm$^3$/g was achieved when reinforcing the *Chlorella* with 10 wt.% HEC, while the specific bending strength of the material with 3 wt.% HEC was 0.40 MPa・cm$^3$/g (Fig 3F). We suggest that the improved mechanical properties of the biocomposite arise from two key mechanisms: (i) HEC binding and (ii) reinforcing. Higher HEC concentrations reduce defects due to HEC's binding effect, resulting in a uniform microstructure (Fig. 2J) and increased material density (Fig. 2I). Additionally, the dispersion and bonding of HEC molecules to *Chlorella's* cell walls enabled effective stress transfer between the microalgae cells, allowing the biocomposite to resist higher loads. These mechanisms collectively enhance both the stiffness and strength of the biocomposite.

Our biocomposites exhibited superior mechanical properties compared to those reported for bulk, three-dimensional composite biomaterials designed through bottom-up methods involving eukaryotic organisms, including other microalgae studies[47], plant cells[65], mycelium[66,67], and yeast matrices[68,69]. In these biocomposites, microorganism cells function as the essential building blocks of the material, reaching values of 160 MPa and 17 MPa for stiffness and strength, respectively. Nevertheless, it is essential to note that mycelium-based biocomposites have lower densities than our materials[66,70].

Analyzing the beam's response to bending loads provides valuable insights into how compression and tensile stresses affect the biocomposite material behavior (Fig. S7a). Using a biocomposite reinforced with 10 wt.% HEC as a model material, as a first order approximation, we employed beam theory to calculate the stress distribution of the structure. The maximum stresses calculated at the top and bottom surfaces of the beam (compression and tension surfaces, respectively) were ~15 MPa (Fig. S7b). This stress level falls below the material's maximum compressive strength of 21 MPa, indicating structural integrity under compressive loads (Fig. 3C). Conversely, although specific tensile strength data were not measured, the bending test revealed crack initiation from the bottom surface of the beam (Fig. S7c), suggesting potential weakness of *Chlorella* biocomposites under tension.

**3.5 Moisture content tunes the stiffness of *Chlorella* biocomposites, which is limited by the HEC $T_g$**

Moisture and temperature are critical environmental factors that significantly affect the mechanical performance and structural stability of *Chlorella* biocomposites. Elevated moisture levels induce HEC gelation (Fig. 2K), leading to a softening of the biocomposite material and a decrease in its stiffness. Moreover, shrinkage can lead to geometric distortions, developing stress concentrations that can result in a weaker structure. Conversely, high temperatures can cause cell wall failure in *Chlorella* and contribute to protein damage within the cell wall structure, potentially compromising the stiffness of the biocomposite[71]. To assess these effects, we conducted 3-point bending tests under various dehydration stages and temperature conditions to analyze how moisture content and temperature variations impact the biocomposite stiffness. Potential weakening effects from cell wall hydration are not considered in the discussion, as the investigation of the hygroscopic properties of dead and alive *Chlorella* cells was not within the scope of this work.

The volumetric shrinkage ratio (VSR) of the biocomposite, observed over different dehydration periods, shows that the most significant shrinkage occurs within the first two days post-3D printing (Fig. 3G), as discussed in Section 3.2 and Supplementary Note #2. During the first stage of dehydration, the CRP, the material undergoes a substantial 50% volume reduction, achieving its final dimensions after 4 days, resulting in an overall volumetric shrinkage of 60%.

We performed 3-point bending tests on biocomposites reinforced with 10 wt.% HEC every day during the dehydration process, to understand the effect of internal moisture on the mechanical response (Fig. 3H). The structures did not fail catastrophically under bending stress at moisture levels above 14%. Figure 3I demonstrates that reducing moisture content from 21% to 6% consistently increases the stiffness of the biocomposite, from 20±9 MPa to 1310±230 MPa, respectively. This behavior can be explained because *Chlorella* cells and HEC molecules, with hydrophilic functional groups like hydroxyl (–OH), are highly sensitive to humidity. These groups preferentially form hydrogen bonds with water molecules instead of adjacent proteins or polysaccharides, weakening the interactions between components and compromising the mechanical properties of the biocomposite[72,73].

The effect of temperature on the mechanical properties was studied using dynamic mechanical analysis (DMA) under 3-point bending. Like the results from the quasi-static bending test (Fig. 3D-E), the storage modulus ($E'$) and loss modulus ($E''$) of the *Chlorella*-based biocomposites show increased values with higher HEC concentrations (Fig. 3J-K). This indicates that HEC reinforcement enhances both the stiffness and energy dissipation capabilities of the material. After the thermal evolution between 25°C and 103°C, the biocomposite stiffness drops significantly from 1.8 GPa to 0.12 GPa, with 10 wt.% HEC reinforcement (Fig. 3J), reflecting the behavior of the *Chlorella* cells at high temperatures[41]. The reduction in the loss modulus (Fig. 3K) suggests reduced internal friction between the biocomposite components due to the softening of the HEC phase, resulting in less energy being dissipated as heat. The improved energy dissipation is further evidenced by the higher peak in the *tan δ* values shown in Fig. 3L, which indicates a transition from a glassy to a rubbery state at 103°C.

The glass transition temperature ($T_g$) of HEC has been reported to be between 106°C and 120°C, depending on the instrument used for measurement [74,75]. The *tan δ* plot (Fig. 3L) suggests that the $T_g$ of HEC in the biocomposite is 103°C. While $T_g$ is often identified from the peak in the loss modulus ($E''$), which represents maximum energy dissipation, the peak was not prominent in our analysis (Fig. 3K). This could be due to several factors such as the relatively low concentration of HEC within the biocomposite or the interactions between *Chlorella* and HEC, which can affect the sensitivity of the measurement. Despite this, the *tan δ* peak remains a reliable indicator of $T_g$ as it represents the ratio of the loss modulus to the storage modulus and provides a clear transition point in the viscoelastic behavior of the biocomposite.

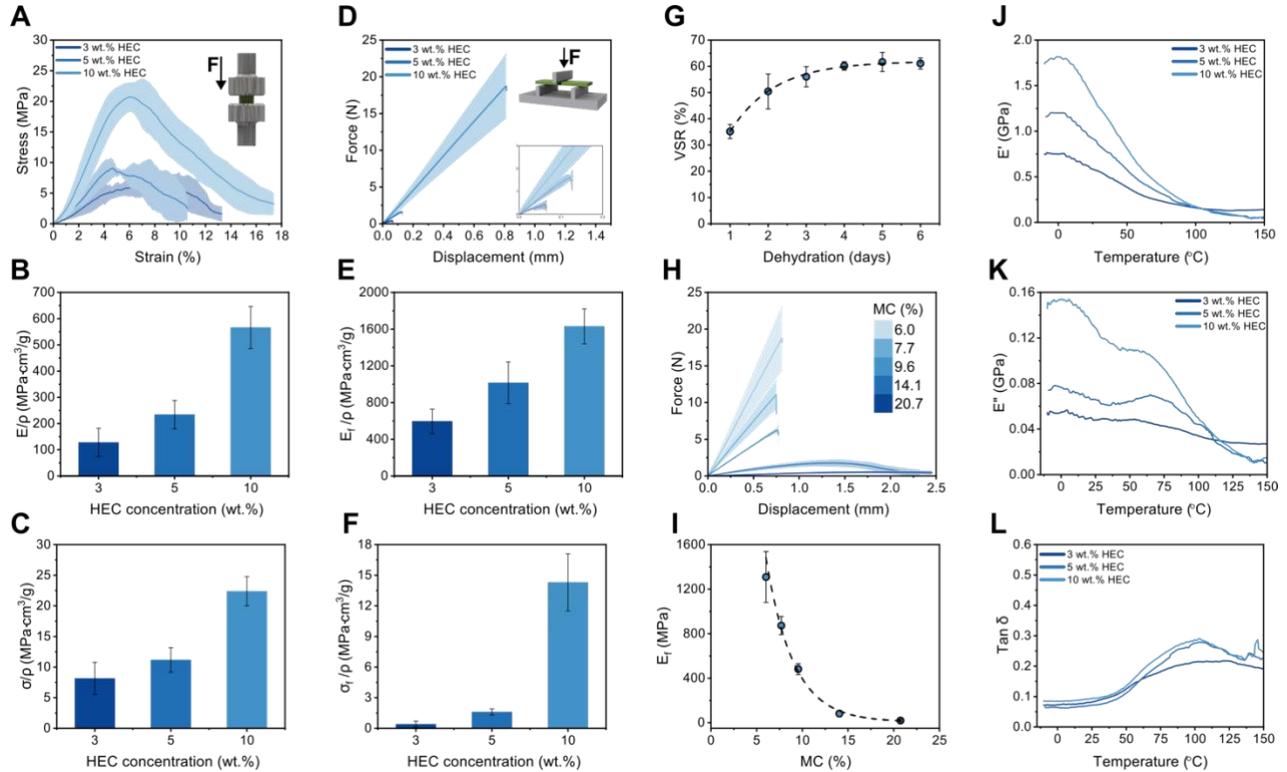

**Fig. 3 Mechanical characterization of the biocomposites.** (A) Average stress-strain and (D) force-displacement curves under compression and 3-point bending, respectively, of the biocomposites with varying HEC levels. The modulus and strength of the biocomposite material increases with HEC concentration under (B-C) compression and (E-F) bending stress. (G) The characteristic volumetric shrinkage of the samples at different dehydration times. (H) The average force-displacement curves of 10 wt.% HEC reinforced *Chlorella* were measured in 3-point bending as a function of moisture content during the dehydration process. (I) Bending modulus of 10 wt.% HEC reinforced *Chlorella* at different moisture contents corresponding to their respective dehydration time. The storage (J) and loss (K) moduli of the *Chlorella* biocomposites decreases with temperature. (L) *Tan δ* of the *Chlorella* biocomposites as a function of temperature.

### 3.6 Lightweight *Chlorella* biocomposites achieve isotropic thermal insulation and mechanical stability

Efficient energy and heat dissipation are characteristic properties of biomass-based materials[76–78]. *Chlorella* biocomposites are expected to also exhibit good thermal-management capabilities, thanks to their hierarchical structure and biopolymer-based composition[79]. We investigated the thermal insulation capabilities of *Chlorella* biocomposites and their resistance to heat flow, measuring the 1-D thermal conductivity using a steady-state method (Fig. S8a). Subsequently, we employed a modified laser-flash method to irradiate the biocomposite surface and explore heat flow behavior in 2D at room temperature (Fig. S8b). Finally, we conducted a 3D simulation to analyze the heat flow behavior. The experiments were performed on samples with a continuous morphology, specifically *Chlorella* reinforced with 10 wt.% HEC.

The thermal conductivity ($k$) of *Chlorella* biocomposites increases with temperature (Fig. 4a). This trend is consistent across different heat fluxes and aligns with the thermal conductivity behavior of bulk polymers[80]. Under extreme conditions at -26.1 ℃, the thermal conductivity values were 0.075 W/mK and 0.095 W/mK for heat fluxes of 0.045 W and 0.21 W, respectively. When the temperature was increased to room temperature and then continuously to 50 ℃, the thermal conductivity for a heat flux of 0.045 W increased from 0.103 W/mK to 0.118 W/mK. For a higher heat flux of 0.21 W, the thermal conductivity of the *Chlorella* biocomposite increased to 0.127 W/mK at room temperature and reached 0.152 W/mK at 50 ℃.

Figures 4B and S11 illustrate the temporal evolution of field temperatures on the irradiated and opposite surfaces of the biocomposite, respectively, showing the nature of two-dimensional heat flow in the sample. These field temperature maps reveal uniform heat dissipation in all directions, consistent with the temperature profiles over time shown in Fig. S10a. At 30 s, the irradiated surface reached a maximum temperature of 41.3 ℃, while the opposite surface registered 26.4 ℃. To analyze heat propagation across the sample thickness, we correlated temperature profiles (Fig. S10) and field temperatures (Fig. 4B and S11) with a 3D model. The agreement between experimental 2D results and 3D model was validated using thermal conductivity data from steady-state experiments and heat capacity calculations from modulated DSC (Fig. S12), all of which demonstrated the sample's heat conduction to be isotropic in nature. The resulting 3D temperature fields illustrate isotropic heat flow over different time intervals.

Thermal conductivity in polymeric materials is influenced by several factors such as crystallinity, processing conditions, morphology, defects, and thermal anisotropy[81]. The heat transfer of *Chlorella* was not affected by the infill pattern from the 3D printing process, as is often observed in 3D printed bulk polymers and their composites[82]. Bulk polymers, characterized by high molecular weight and random entanglements with low degrees of crystallinity and large free volumes, typically exhibit thermal insulation properties as a result of phonon scattering. In contrast, the *Chlorella* biocomposite developed in this study demonstrates superior insulation compared to bulk polymers, partly due to its low density of 0.96 g/cm³. To illustrate this, thermal conductivity at room temperature was plotted against density for various bulk polymers and biomass-based materials (Fig. 4C). Biomass-based materials, like wood and lignocellulosic fibers, with densities up to 0.75 g/cm³, benefit from their porous structure, resulting in thermal conductivity values below 0.1 W/mK. Conversely, materials with continuous morphologies and higher densities (>1 g/cm³), such as HDPE with a thermal conductivity up to 0.45 W/mK, exhibit higher thermal conductivity values by improving phonon transport and reduced air-material interfaces.

Wood materials often enhance thermal insulation through delignification, which not only removes lignin but also introduces nanopores within cell walls[79]. This process reduces phonon transport while enhancing anisotropic thermal conductivity due to improved alignment of cellulose nanofibrils. In contrast to wood and natural fibers, *Chlorella* microalgae lacks the presence of lignin within the cell composition[49]. Moreover, *Chlorella* biocomposites demonstrate isotropic heat transfer while maintaining a continuous morphology, offering mechanical properties comparable to both bulk polymers and wood, alongside effective thermal insulation (Fig. 4D).

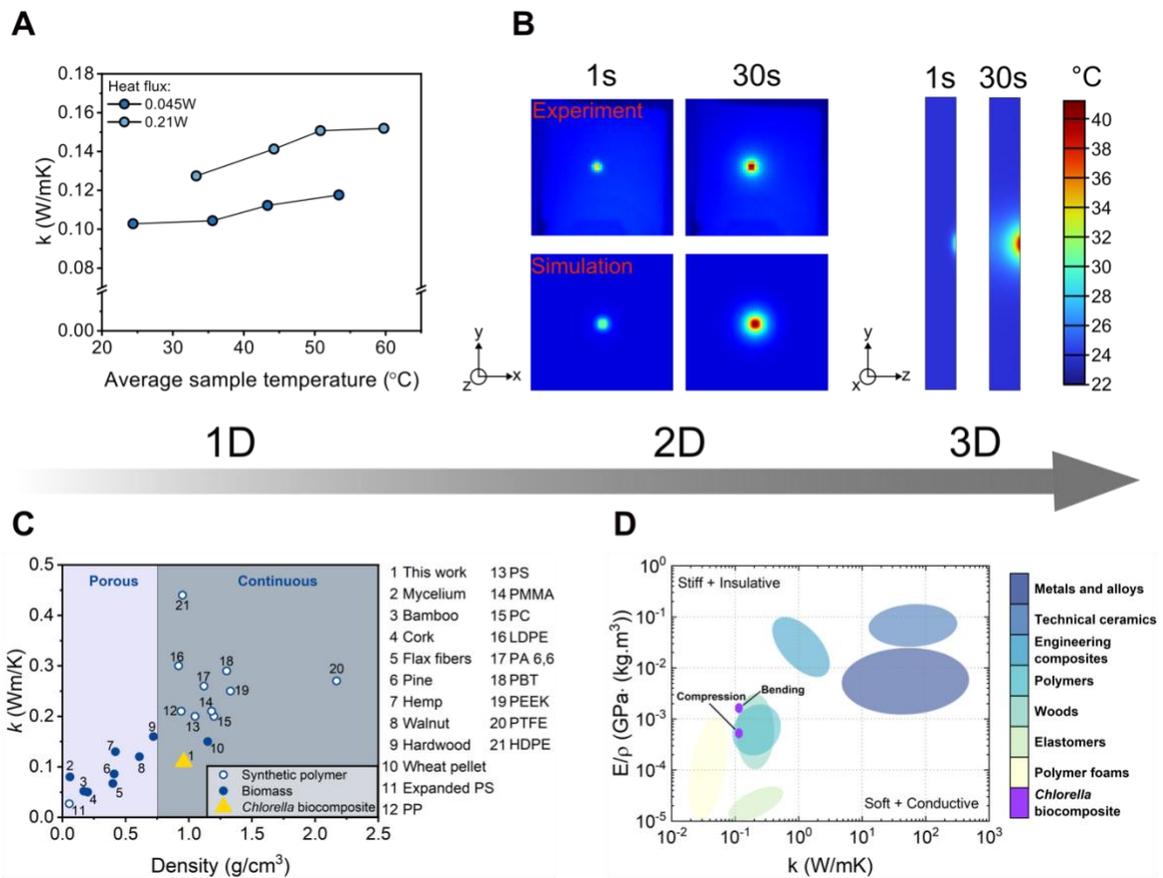

**Fig. 4 Multidimensional thermal conductivity of *Chlorella* biocomposite.** (A) Experimental results of 1D thermal conductivity test at different ambient temperatures and heat fluxes. (B) Experimental and simulation correlation of the propagation of temperature fields in different time steps after laser emission on the irradiated surface, and simulation of 3D heat flow through the thickness of the biocomposite. (C) Comparison of thermal conductivity, as a function of density, of *Chlorella* biocomposite (yellow triangle) and diverse biomass[76–78,83–86] (white filled circles) and synthetic polymeric[81,87,88] (blue filled circles) materials, with an emphasis on their morphology. (D) Normalized Young's modulus vs thermal conductivity for various materials and our biocomposite (in purple) under compression and bending tests, respectively.

## 4. Summary and conclusions

We provided a systematic approach for the development of a natural and biodegradable hierarchical biocomposite material through extrusion 3D printing. Our study demonstrates that *Chlorella*-based bioinks can withstand high shear forces during the 3D printing process, effectively preserving the integrity of their cellular structure. The rapid viscosity recovery mechanism enables the 3D-printed *Chlorella* materials to maintain their shape post-printing. A practical dehydration protocol was developed, resulting in crack-free biocomposite materials with continuous morphologies of aggregated cells. Water significantly influences the rheological properties of the bioink without adversely affecting the mechanical properties of the biocomposites. This finding allows for optimal printability while maintaining mechanical performance. Bioink and process optimization were implemented to maximize structural capabilities of the biocomposite. HEC served a dual role as both a binder and reinforcement, significantly enhancing the compression and bending moduli of the composite material to 566 MPa and 1630 MPa, respectively. Further reinforcement can be achieved by integrating load-bearing components, such as natural fibers or nanoclay. The *Chlorella* biocomposites demonstrated isotropic heat transfer and served effectively as thermal insulators, with a thermal conductivity value of 0.10 W/mK at room temperature. The results presented in this study are specific to the natural model *Chlorella vulgaris* microalgae. We anticipate that these observations can be extended to other microalgae strains with different cell shapes and sizes. Additionally, similar studies on microalgae from various sources, such as the wastewater industry, may provide a sustainable alternative to the current materials global demand. Further research is expected to solidify our findings and address several key areas: (1) the capability of *Chlorella* biocomposites to resist impact loads through fracture toughness; (2) the effect of post-printing treatments such as annealing on biocomposite morphology and mechanical properties; (3) the enhancement of stiffness and thermal insulation capabilities by introducing natural fibers into the biocomposite system; and (4) the potential for large-scale fabrication of intricate 3D structures. *Chlorella* biocomposites hold promise as potential substitutes for bulk polymers and wood in applications requiring both stiffness and thermal insulation.

**Supporting information**

Supporting information is available from the Wiley Online Library or from the authors.

**Acknowledgments**

CD and IK acknowledge the NSF for their financial support through the grant no. 2308575. This work was supported in part by the Resnick Sustainability Institute at Caltech. The authors gratefully acknowledge the support and infrastructure provided for this work by The Kavli Nanoscience Institute at Caltech. IK acknowledges Fulbright Israel for the financial support. DT and MB acknowledge financial support from the Harvard University Center for Green Buildings and Cities and the Joint Institute for Housing Studies. The authors acknowledge Prof. Sergio Pellegrino, and the technical support provided by Dr. Rachel Behrens on the DMA tests.

**Conflict of interest**

The authors declare no conflict of interest.

**Data availability request**

The data that support the findings of this research are available from the corresponding author upon reasonable request.

**Supporting Information**

1. Experimental section.
2. Supplementary figures: Fig. S1 to Fig. S12.
3. Supplementary notes 1 and 2.
4. Supplementary references.

**Experimental**

*Materials*

All materials were used as received unless otherwise specified. Ultra-pure *Chlorella vulgaris* microalga (Fig. S1), in dried powder form, was obtained from eChlorial (L'Isle-d'Abeau, France). This freshwater microalga is cultivated in glass tubes under natural light conditions, reducing the abundance of contaminants of the final specimens that could interfere with a comprehensive scientific assessment of the microalgae as a matrix composite. 2-Hydroxyethyl cellulose (HEC) ($M_v \sim$ 1,300,000) and Magnesium nitrate hexahydrate were purchased from Sigma-Aldrich (MKE, USA). Osmium tetroxide ($OsO_4$) 4% aqueous solution was obtained from Electron Microscopy Sciences (PA, USA).

*Biocomposite ink preparation*

Different bioink compositions were synthesized for the 3D printing of the structural geometries. *Chlorella*-based bioinks were prepared in glass beakers by initially dispersing HEC in DI water, acting here as the solvent in the bioink. Then, the microalgae powder was added to the mixture, and the bioink was manually mixed until a homogeneous paste was obtained. Water concentration (58 wt.% unless otherwise specified) is relative to the total weight of the bioink, whereas the concentration of HEC (3, 5, and 10 wt.%) is relative to the mass of *Chlorella* only. The water concentration is relative to the bioink total weight, whereas HEC concentration is relative to the solid biomass microalgae concentration.

*3D Bioprinting of hierarchical biocomposites*

The different biocomposite inks were loaded into 10 ml disposable plastic syringes and centrifuged for 3 minutes at 3500 RPM (5804, Eppendorf) to remove air bubbles and evenly set the bioink on the bottom of the syringe. After centrifugation, a syringe rubber cap was inserted into the top end of the syringe and a needle was used to remove the air gap in between. All structures were 3D printed in a Allevi 2 Bioprinter (3D Systems, PA, USA). The print paths for the 3D printed structures were designed using Grasshopper (Rhinoceros, Robert McNeel & Associates) and exported as a G-code format for further printing. The custom code enabled a continuous extrusion of the bioink filament to avoid discontinuities along the printed structures, thus overcoming the disadvantage of extrusion 3D printing viscous materials in pneumatic-based systems. Unless otherwise stated, parts were printed using tapered tips with an inner diameter of 1.19 mm (Nordson, Westlake, OH, USA) onto a film substrate made of fiber glass reinforced polytetrafluoroethylene with a thickness of ~25.4 $\mu$m (McMaster-Carr, IL, USA). Printing parameters included pressures between 1.4 and 7.6 bar and speeds between 3 and 15 mm/s. Structures were printed with a layer height of 1 mm. To reduce natural shrinkage and anisotropic effects because of printing orientation, structures were printed with a 0-90° infill pattern with a 1.16 mm infill distance. Samples of 20 x 20 x 10 mm$^3$ and 70$l$ x 13$w$ x 4$t$ mm$^3$ for compression and bending test were printed, respectively. After printing, structures were peeled off the substrate, turn onto their side, exposing their larger surface area towards drying, and incubated (Versatile Environmental Test Chamber, MLR-352H, Panasonic) for 2 days at a relative humidity of 75%. Then, a second incubation period of 4 days, at ~RH=50%, took place in a humidity chamber controlled by a saturated solution of magnesium nitrate (200 gr: 30 ml)[1] to complete the drying process (supplementary note #2). After those 6 days, the samples weight remained steady, and the structures were stored in a desiccator (RH~20%) until testing. The relative humidity was tracked using a humidity and temperature smart sensor HT1 sensor (SensorPush, Cousins and Sears Creative Technology, NY, USA) with a humidity accuracy of ±3%. Before testing, samples were dried in an oven (Thermoscientific) at 60℃ overnight to remove any moisture residues within the structure.

*Rheological analysis of bioinks*

To evaluate the suitability of the bioinks to the 3D printing process, the rheological properties of the biocomposite inks were characterized using a dynamic rheometer (Discovery HR-20 hybrid rheometer, TA Instruments, Delaware, USA). The instrument was equipped with a 40 mm parallel plate geometry, and the tests were performed using a gap height of 1 mm and a solvent trap to prevent solvent evaporation during the test. All measurements were completed at 20°C. Flow sweeps were performed to study the shear thinning behavior at shear rates ($\dot{\gamma}$) between 10$^{-2}$ and 10$^2$ s$^{-1}$. Oscillation amplitude sweeps from 10$^{-3}$ to 10 kPa, with an induced shear stress at 1 Hz, were performed to define the linear viscoelastic regime (LVR). Yield stress ($\tau_y$) was reported by recording the corresponding value of oscillatory stress at the crossover points of storage (G') and loss (G") moduli. For the viscosity recovery tests, the viscosity ($\eta$) of the bioinks was measured for 100 s, while the shear rate was adapted to 10$^{-2}$ s$^{-1}$ for 30 s in stage I, to $10^2$ s$^{-1}$ for 10 s in stage II, and finally, for $10^{-2}$ s$^{-1}$ for 60 s in stage III. This method was selected to simulate the printing process and to evaluate the recovery of the viscosity of the algae-based biocomposites. Before collecting the data, the instrument was allowed to stabilize the shear rate for an initial period of 30 s (Fig. 2H).

*Scanning Electron Microscopy (SEM)*

The preparation process for the structural analysis of the specimens began by exposing the surface of interest via cryogenic failure with liquid nitrogen. Then, samples were fixed and chemically stained by evaporating a solution of 4% $OsO_4$ at room temperature for 1 h directly to the surface of interest. Besides fixing the biological sample, and additional benefit of using osmium is its high atomic number (76) that enhances the image contrast in the electron microscope[2]. To evaluate the micro-morphology of the 3D printed materials, high-resolution scanning electron microscopy (HRSEM) images were taken using a Nova600 NanoLab (FEI, now ThermoFisher). Images were obtained using a secondary electron (SE2) detector at a working distance of 6-7 mm and an acceleration voltage of 5-10kV. Prior to SEM imaging, the samples were coated with a 5nm layer of platinum using a high-resolution sputter coater (Cressington 208HR).

*Optical imaging*

Optical images were taken with a camara tube (Optem Fusion Micro-imaging system, Qioptiq) on the cross-section of printed structures and the surface of filaments were taken to measure the filament diameter and to evaluate the morphology of the printed samples.

*Mechanical analysis*

Compression and bending tests were carried out to characterize the mechanical properties of the different biocomposites at a strain rate of $10^{-3}$ s$^{-1}$. Quasi-static uniaxial compression tests of at least five cubic samples (~16 x ~16 x ~7 mm$^3$) were performed on a Instron 5560 universal system (Instron, MA, USA). The instrument was equipped with a 50 kN load cell and the test strain was monitored using a laser extensometer, model LE-05 (Electronic instrument research, PA, USA), with a scanning rate of 100 scans/s. Before testing, samples were polished to ensure an even surface area.

Quasi-static three-point bending tests of at least five printed rectangular beam-shaped samples (~51.0*l* x ~9.5*w* x ~3.0*t* mm$^3$) were carried out using an Instron E3000 instrument (Instron, MA, USA) equipped with a loadcell of 250 N. The test span was 46 mm, resulting in a specimen span length to thickness ratio of ~17. From each load-displacement curve, the flexural modulus $E_f$ (MPa) and bending strength $\sigma_f$ (MPa) were calculated according to the following formulae:

$$E_f = \frac{m}{4w}\left(\frac{l}{t}\right)^3 \qquad (E1)$$

$$\sigma_f = \frac{3Pl}{2wt^2} \qquad (E2)$$

where $m$ (N mm$^{-1}$) is the slope in the elastic region of the force-displacement curve, $l$ (mm) is the span length, $P$ (N) is the applied load, $t$ (mm) is the specimen height, and $w$ (mm) is the specimen width. To characterize the effect of moisture content on the bending modulus of the *Chlorella*-based biocomposites, at least five 3D printed structures were tested at different stages of dehydration within the drying period. For each set of samples, the moisture content MC (%) and volumetric shrinkage ratio VSR (%) were calculated using the following formulae:

$$MC = \frac{W_w - W_c}{W_x} \cdot 100 \qquad (E3)$$

$$VSR = \frac{V_i - V_f}{V_i} \cdot 100 \qquad (E4)$$

where $W_w$ and $V_i$ are the weight and volume of the wet sample, or as printed, respectively. $W_c$ and $V_f$ are the weight and volume of the sample prior to testing, respectively. We consider the tested specimen to have equal dimensions and load-displacement curve to a beam that is isotropic and homogeneous, and therefore, the flexural modulus and strength of the tested specimen can be roughly comparable to the tensile modulus and strength, respectively. All data analysis was performed using OriginPro 2022.

*Dynamic Mechanical Analysis (DMA)*

A dynamic mechanical analyzer (Discovery DMA 850, TA Instruments, Delaware, USA) was used to measure the effect of temperature on the bending moduli of the *Chlorella*-based biocomposites. The tests were conducted in a three-point bending mode with a support span distance of 20 mm. The sample's width and thickness were similar to those used for the quasi-static three-point bending test (~9.5$w$ x ~3.0$t$ mm$^2$). The composites were tested from -10 ℃ to 150 °C with a heating rate of 3 °C/min at an oscillation frequency of 1 Hz. The oscillation amplitude of the test was kept at 10 $\mu$m.

*Modulated Differential Scanning Calorimetry (MDSC)*

The heat capacity ($C_p$) across different temperatures (-30 °C to 90 °C) of *Chlorella* reinforced with 10 wt.% HEC was calculated using a DSC 25 (TA Instruments, Delaware, USA). The instrument, in a modulated mode, was calibrated using a sapphire sample under air conditions. The tests were conducted at an amplitude of ±1 °C, for 120 s, and with a heating rate of 2 °C /min. Five samples were tested with masses ranging between 11.5 and 13 mg.

*Thermal conductivity test*

The thermal conductivity (*k*) of *Chlorella* reinforced with 10 wt.% HEC was measured at ambient temperatures between -30 and 50°C using a 1D steady-state conduction guarded heat flow method. The tested sample had a cross-sectional area of 400 mm$^2$ and a thickness of 2.1 mm. A schematic representation of the setup is illustrated in Fig. S8a. The test format consisted of a sandwich-like configuration, where the chlorella biocomposite was placed between a 25.4x25.4 mm$^2$ adhesive back-heater patch (McMaster, model 35475K242) and an aluminum block, which performed as a heat sink guaranteeing 1D heat flow. Thermocouples (type E) were placed at the interfaces between the sample and the adhesive patch and the sample and the aluminum block, respectively. The temperature from the thermocouples was recorded using an Omega RDXL4SD thermocouple reader. To account for any heat loss due to convection from the environment, we covered the top of the setup with a 15 mm layer of cotton (k=0.035 W/mK[3]). Two different heat fluxes ($\dot{Q}$) were used to test the thermal conductivity of the biocomposite, 0.18 and 0.045 W. Voltage and current supplied to the heater patch were adjusted using a power supply (Tekpower, TP1803D). The thermal conductivity was calculated using the following equation:

$$k = \frac{\dot{Q}L}{A\Delta T} \quad \quad (E5)$$

where *A* and *L* represent the cross-sectional area and thickness of the sample, respectively, and *ΔT* represents the temperature gradient across the sample thickness as measured by the thermocouples.

For the laser flash analysis on the same sample dimension as the steady-state test, a Melles Griot laser (Melles Griot, CA, USA) with a power of approximately 16 mW was used to irradiate the sample surface at with a spot size of 3 mm (Fig. S8b). An infrared camera (FLIR SC6000) monitored the temperature fields on both the irradiated and opposite surfaces of the sample in real time in consecutive trials. Before testing the chlorella sample, the setup was calibrated using a Kapton sample of the same dimensions. Data for the steady-state experiment was reordered for approximately one minute after the system reached a steady-state temperature, while that of the laser flash experiments was recorded for thirty seconds after the start of heating, long enough to reach steady state (Fig. S9).

*Finite element method*

COMSOL Multiphysics software was used to simulate and illustrate the 3D heat transfer of *chlorella* biocomposites. Justified by our steady-state and laser flash experiments, isotropic thermal conductivity was used and set to a value of .12 W/m K. The heat capacity of the material was taken from the modulated DSC experiments (Fig. S12) and was set at 1700 J/kg K. The sample was modeled using mapped solid elements (<.2 mm) with these isotropic thermal properties (Heat Transfer in Solids module). Heat loss via convection was modeled by applying an outward heat flux on the front of the sample. Laser heating was implemented using a point illumination source (Surface-to-Surface Radiation module, ray-shooting radiation method) and a nonphysical mask covering the sample, one with a central hole of the same dimension as the experimental laser spot; this arrangement simulated both the finite laser spot size and nonuniformity in the spot heating profile itself. The simulated laser source power and convective heat loss (both unknown) were then adjusted to match the 2D experimental data on front and back of the sample based on the known experimental isotropic thermal conductivity. From an initial room temperature (23.6ºC), the model was simulated using a 30s transient thermal analysis with a room temperature ambient radiation boundary condition.

Statistical analysis

Quantitative results are expressed as mean ± standard deviation (SD). One-way analysis of variance (ANOVA) with posthoc Tukey's test using OriginPro 2022 software was applied to determine whether significant differences existed between the mean values of the experimental groups. A difference between groups was considered to be statistically significant at $p<0.05$ (ns denotes $p>0.05$).

**Supplementary figures**

a)  b)

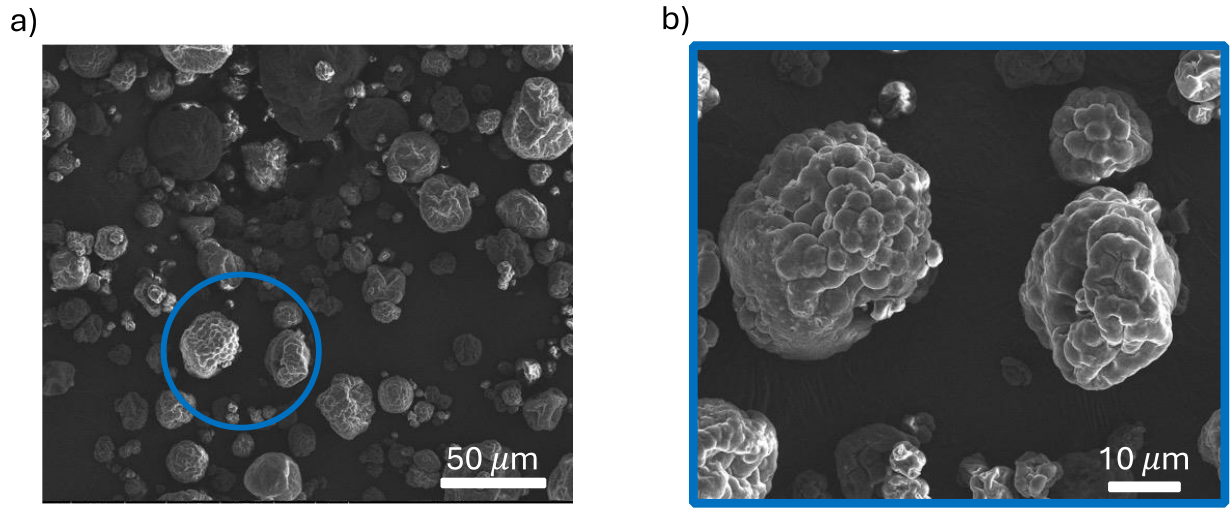

**Fig. S1** a) SEM micrograph of *Chlorella vulgaris* powder before dispersion in water. b) Magnification of the area within the blue circle in a). The images exhibit an agglomeration of *chlorella* cells because of the spray drying process. Individual dried *chlorella* cells have a diameter of 3.21±0.21 $\mu$m.

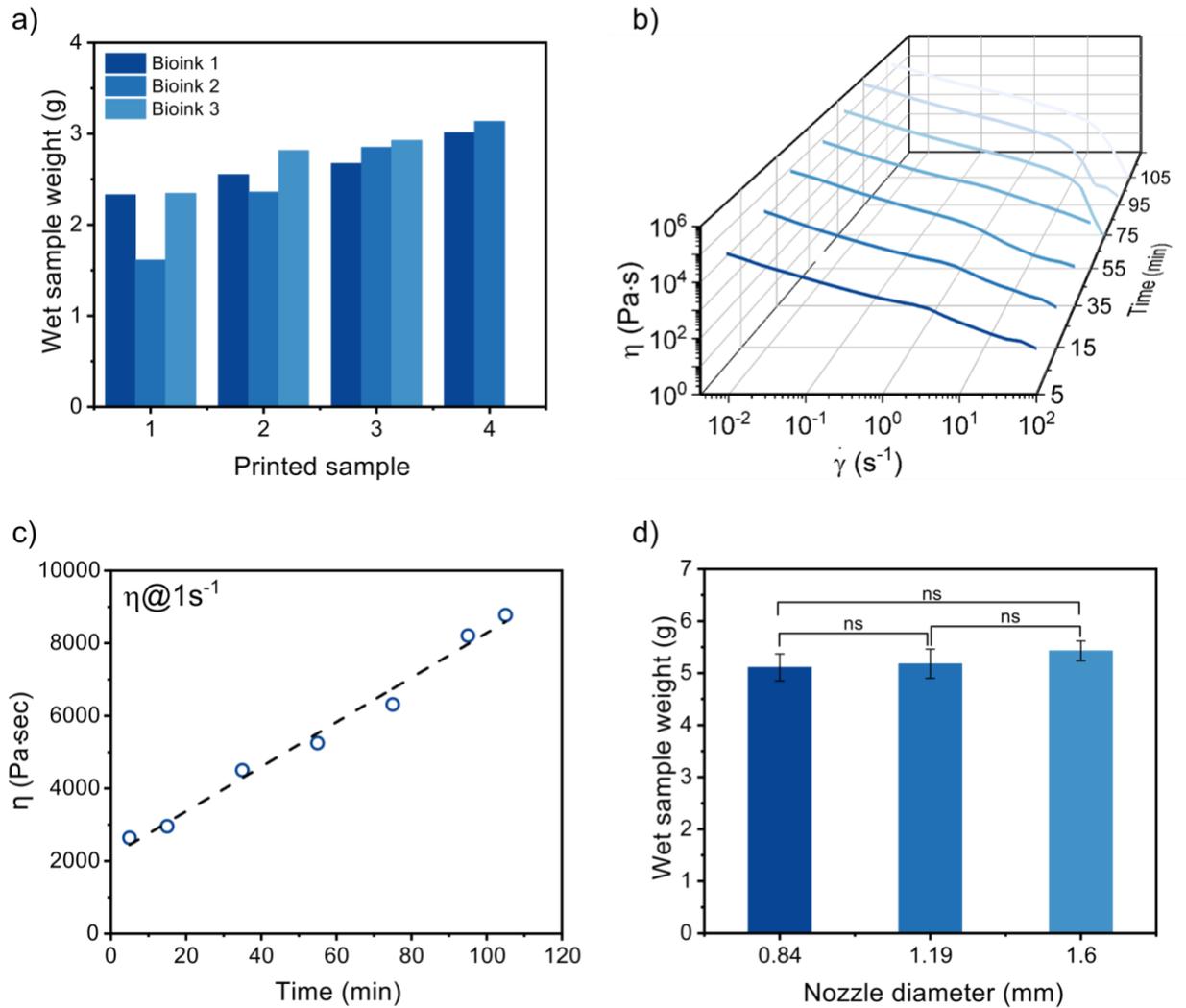

**Fig. S2** a) Varying sample weight after 3D printing. b) Change of bioink viscosity over time. c) Linear increasing in bioink viscosity over time at shear rate of 1 s$^{-1}$, representative of the consistency index (*k*) in the power law model. d) Consistent weight of 3D printed structures after adjusting the printing speed based on the specific mass flow rate of the bioink (supplementary note#2). All plots are for an HEC concentration of 5 wt.%.

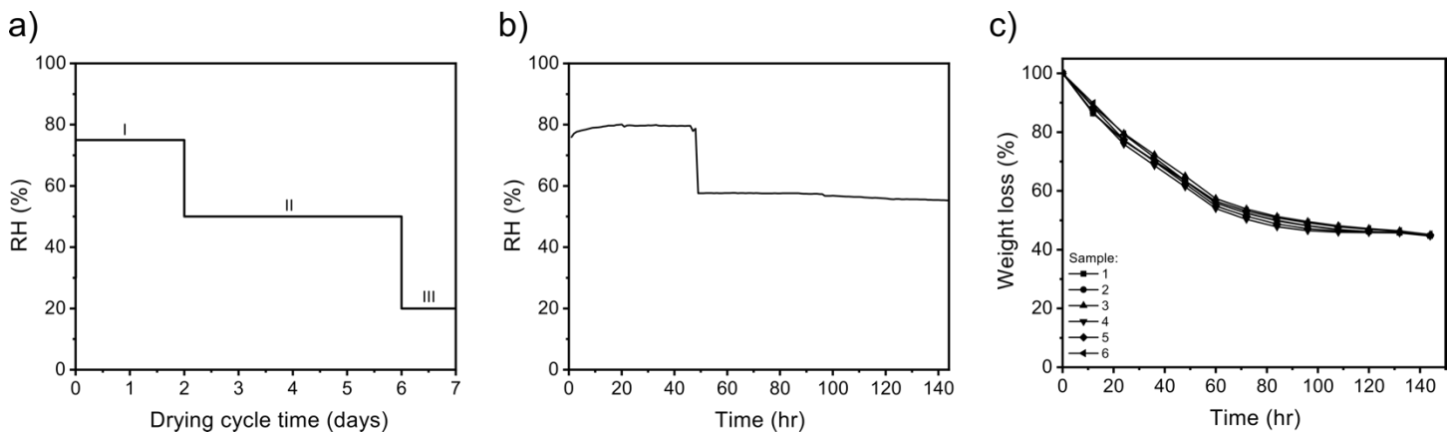

**Fig. S3** a) Stages during the dehydration process; I) constant rate period at RH=75% for 2 days; II) falling rate period at RH=50% for 4 additional days; III) storage at RH=20% until testing. b) Humidity sensor reading during stage I and II showing higher levels of humidity because of water evaporation from the sample. c) Weight loss of the samples during the dehydration process. After 144 hours, all samples lost ~45% of weight as a result of evaporation. Further evaporation occurred at lower RH values during storage before testing.

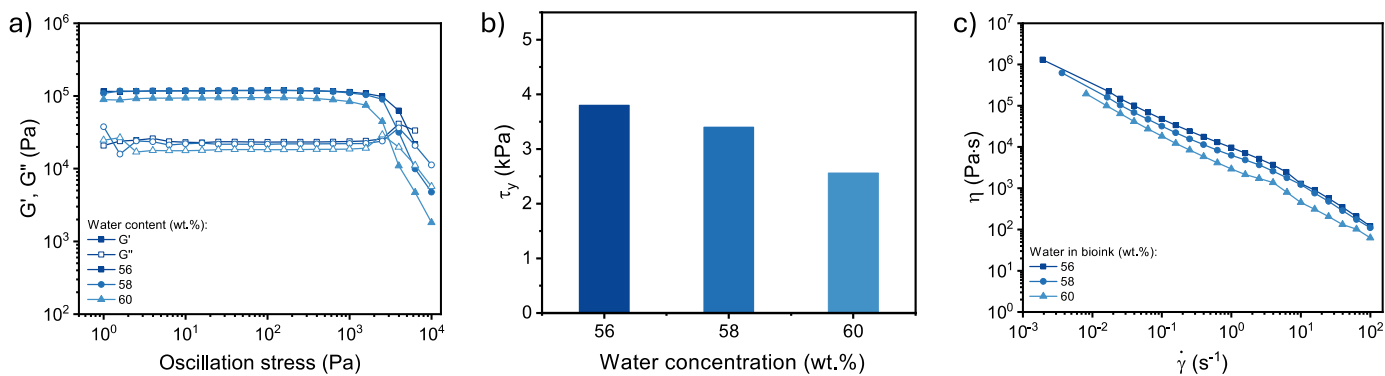

**Fig. S4** a) Oscillatory tests of the different biocomposites. The G' and G'' crossover point represent the shear stress at yield of the biocomposite, plot in b). c) Flow analysis of the biocomposites demonstrating the reduction effect of water on the viscosity of the material.

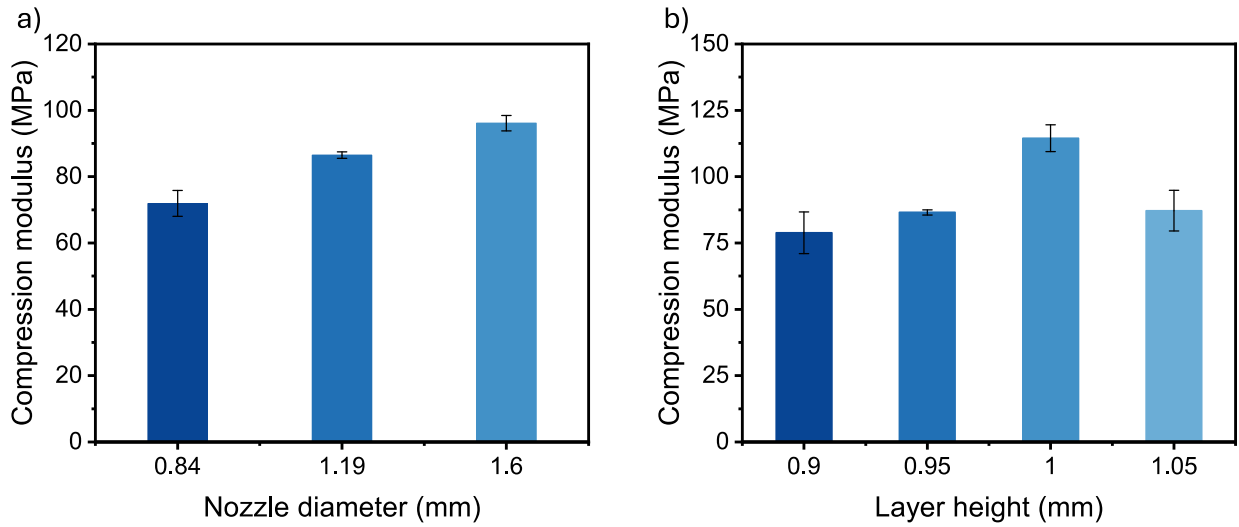

**Fig. S5** Dependence of a) nozzle inner diameter and b) layer height on the compressive modulus of *chlorella* reinforced with 5 wt.% HEC.

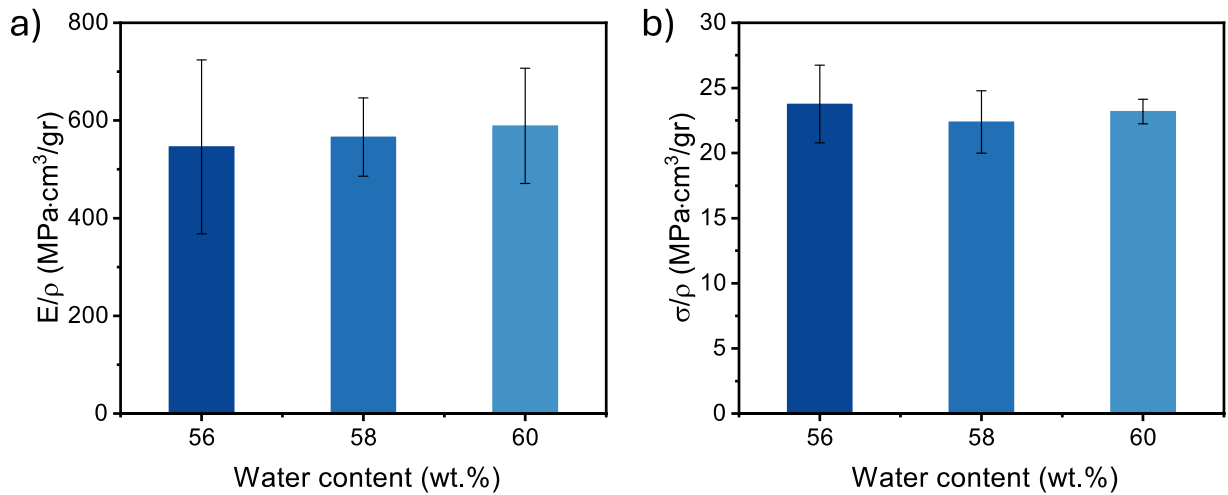

**Fig. S6** a) Normalized compression modulus and b) normalized compression strength of the biocomposites with different water concentrations in the bioink.

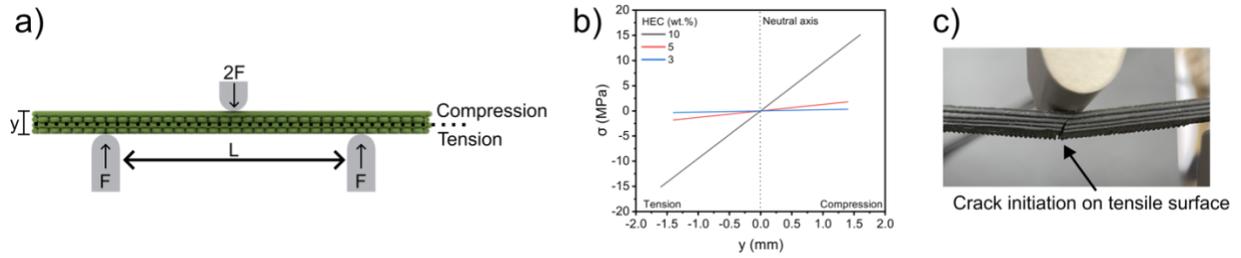

**Fig. S7** Failure behavior under bending. a) Free body diagram of the *chlorella* biocomposite beam under 3-point bending. b) Distribution of stresses across the beam geometry for the different biocomposites. c) Image of the biocomposite under 3-point bending showing the crack initiation at the bottom surface of the beam.

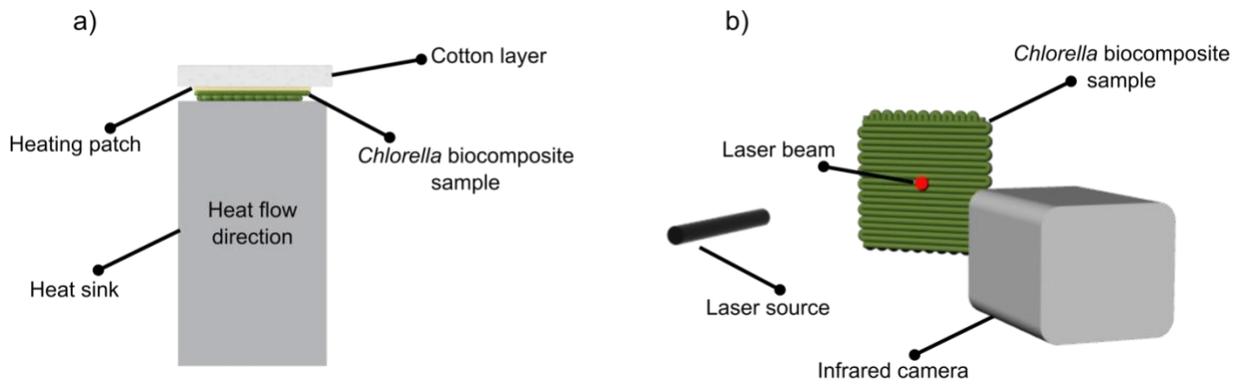

**Fig. S8** Schematic representation of heat transfer experiments. a) Steady state method and b) laser flash method.

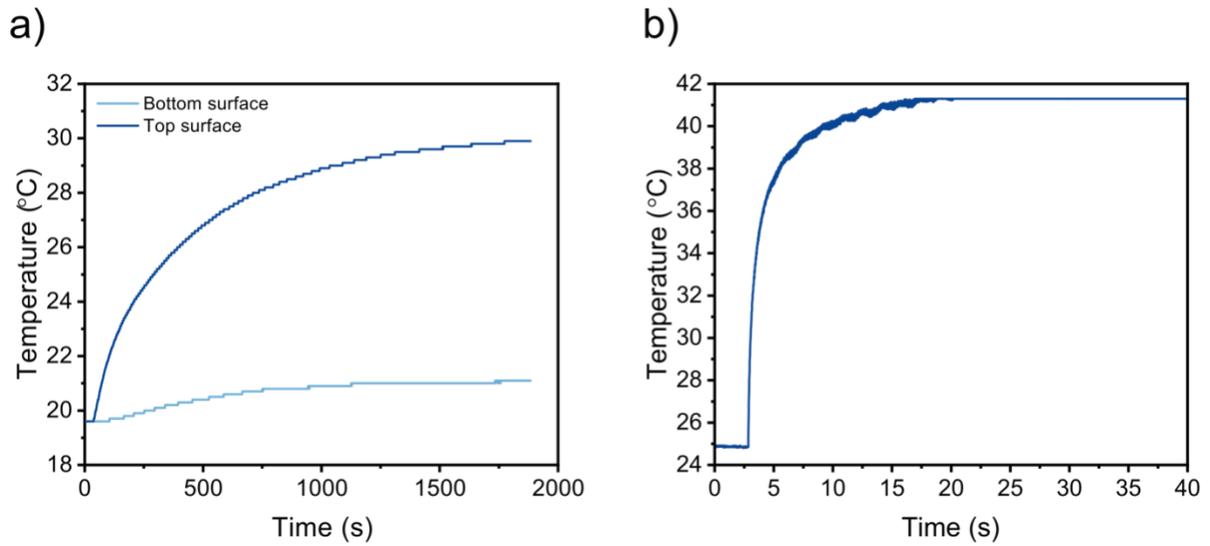

**Fig. S9** Temperature profiles at ambient room temperature for the a) steady-state thermal conductivity measurement and the b) irradiated surface in laser flash experiment. In a), top surface is referred to the sample surface in direct contact with the heating patch.

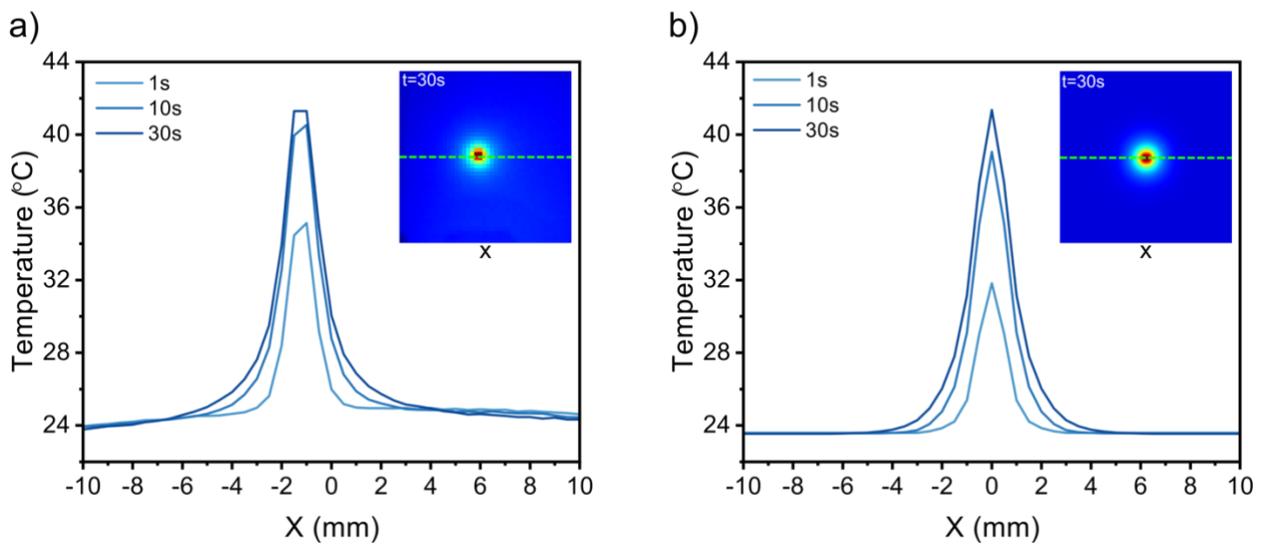

**Fig. S10** Temperature profile on a) irradiated surface of sample at 1s, 10s, and 30s of laser exposure and b) temperature profile correlation from the simulation.

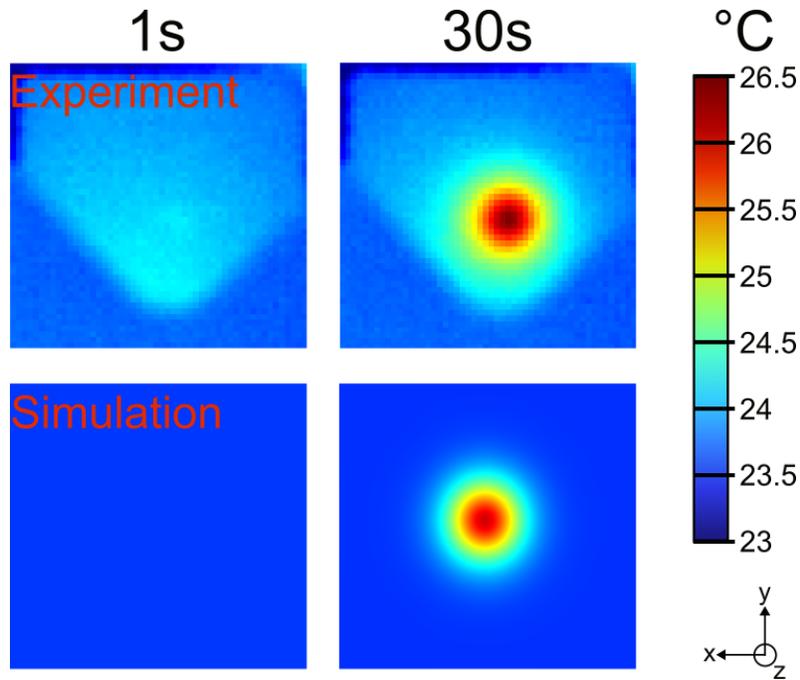

**Fig. S11** Field temperature on *chlorella* biocomposite opposite surface at 1s and 30s of exposure, showing the correlation between the experimental results and the simulation, respectively. V-shaped blue region at the bottom of experiment is part of clamping apparatus.

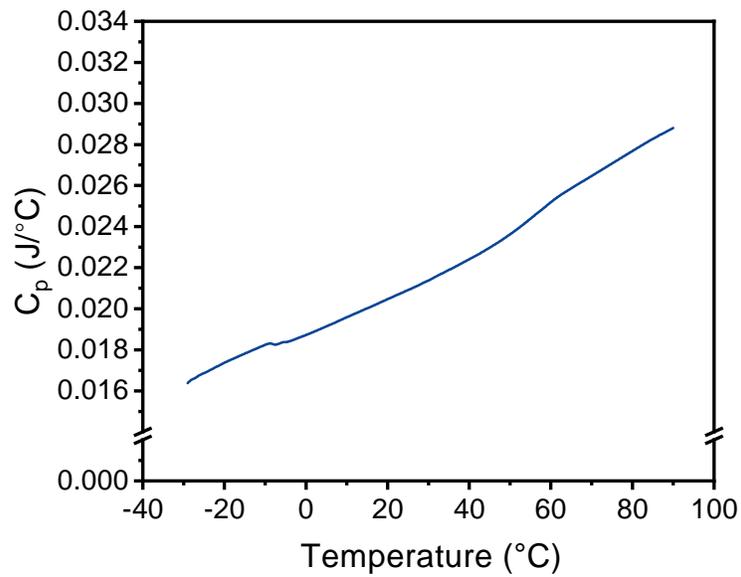

**Fig. S12** Heat capacity of *chlorella* biocomposite reinforced with 10 wt.% HEC.

**Supplementary notes**

Supplementary note #1: Regulating bioink deposition in extrusion 3D printing through the mass flow rate principle to manage viscosity fluctuations:

Consistently maintaining the rheological properties of the bioink during extrusion 3D printing is crucial for ensuring process stability and uniformity in printed components. The bioink, consisting of *chlorella* cells and HEC dispersed in water, exhibits varying viscosity over time due to the hygroscopic nature of its components, particularly HEC, which undergoes a gelation process. Inconsistent bioink viscosity during printing under identical conditions leas to non-uniform material deposition within the printed structures (Fig. S3a-c).

To achieve uniform weight in 3D printed structures, addressing the variability in bioink viscosity over time requires adjusting processing parameters accordingly. Implementing the mass flow rate principle involves dynamically adjusting the printing speed of the bioink based on its current rheological state before each printing operation. This adjustment was calculated using the formula:

$$\dot{m} = v\rho A \qquad (E6)$$

where $\dot{m}$ is the mass flow rate of the bioink, *v* is the printing speed, $\rho$ is the density of the bioink, and *A* is the cross-sectional area at the tip of the nozzle. By calculating the appropriate printing speed for each sample, structures with consistent weight were achieved (Fig. S3d).

Supplementary note #2: Theoretical basis for the development of a dehydration protocol for *chlorella*-based biocomposites:

*Chlorella vulgaris* cells are microspheres that form suspensions when dispersed in water, creating a bioink suitable for extrusion 3D printing. Dehydrating the microalgae-based biocomposite is crucial for achieving high strength and stiffness, as shown in Fig. 3H-I. To prevent structural cracking and failure during drying, a dehydration protocol was developed based on the principles of the sol-gel theory of drying[4]. Understanding the stages of drying, even qualitatively, is critical to develop such a protocol. After 3D printing, the initial condition of the system establishes that there is a solid phase or solid network within a liquid phase.

*Stages of drying:*

During the first stage of drying, known as the constant rate period (CRP), the evaporation rate of water within the biocomposite structure remains relatively constant. At this stage, the volumetric shrinkage of the 3D printed part is proportional to the volume of liquid that evaporates. Evaporation occurs through the diffusion of water from the interior to the surface, driven by differential pressure, creating a liquid film that evaporates. Lower ambient humidity increases the vapor pressure gradient between the capillary pressure, which pulls liquid from the interior, and the ambient pressure, thus enhancing evaporation.

During the CRP, the surface tension of the liquid at the sample surface produces compressive stresses in the solid network due to the pressure gradient between the capillary pressure across the structure thickness and the ambient pressure. This results in differential shrinkage of the solid network. If the tensile strength of the solid network is insufficient to withstand these forces or the differential shrinkage, the structure can crack and eventually fail. To minimize these risks, the developed dehydration protocol increases ambient humidity (Fig. S2). This approach lowers the pressure gradient, drying stresses, and differential shrinkage, reducing the evaporation rate and decreasing liquid transport to the surface of the structure. Increasing ambient humidity is preferred, as the stiffness of the biocomposite material is too low to resist drying stresses (Fig. 3I).

As the surface begins to dry, capillary forces become less efficient at transporting liquid from the interior to the surface, leading to a decrease in the evaporation rate. This stage is known as the falling rate period. During this period, the drying mechanism shifts from being controlled by surface conditions to being controlled by the internal transport properties of the solid network. Most shrinkage occurs during the CRP (Fig. 3G), and at this stage, the structure's stiffness can resist the drying stresses, diminishing shrinkage and reducing the likelihood of cracking and failure. Therefore, the ambient humidity can be lowered to increase the evaporation rate (Fig. S2), as the stiffness of the biocomposite material is sufficient to withstand the remaining internal stresses during the final stage of dehydration (Fig. 3I).

**Supplementary references**